\documentclass[aps,prl,twocolumn,groupedaddress,showpacs,showkeys]{revtex4}
\usepackage{graphicx}
\begin{document}

\title{Possibility of Turbulence from a Post-Navier-Stokes Equation}

\author{Pascal Getreuer}
\affiliation{Mathematics Department\\
University of California, Los Angeles\\
Los Angeles, CA 90095}
\author{A. M. Albano}
\affiliation{Department of Physics \\
Bryn Mawr College  \\
Bryn Mawr, PA 19010}
\author{A. Muriel}
\affiliation{Data Transport Systems\\
347 East 62nd Street\\
New York, NY 10021}
\date{\today}

\begin{abstract}
We introduce corrections to the Navier-Stokes equation arising from
the transitions between molecular states and the injection of external
energy. In the simplest application of the proposed post Navier-Stokes
equation, we find a multi-valued velocity field and the immediate
possibility of velocity reversal, both features of turbulence.
\end{abstract}



\maketitle

\section{Introduction}
Traditionally, any attempt to describe turbulent behavior in fluids
starts with the Navier-Stokes equation (NSE) \cite{Frisch}. However,
success has at best been mixed \cite{Deissler}. In this Letter, we
explore the idea that NSE is not the unique approach to the study of
turbulence, and that turbulence may be found in what we label as
post-Navier-Stokes equations. What could modify NSE?  In our opinion,
the molecular nature of fluids can no longer be ignored
\cite{Nerushev1,Nerushev2,Novopashin1,Novopashin2,Hinkle1,Hinkle2}. So
we attempt to modify NSE by including quantum concepts, and in
particular, in its simplest application, arrive at the possibility of
velocity reversal and multi-valued velocity fields \cite{Sirovich},
both important features of turbulence.  Before showing these final
results, we need to make two comments on the derivation of NSE to
justify our proposed post-NSE.

The traditional way of deriving the Navier-Stokes equation (NSE) is
phenomenological, based on the continuum model and conservation of
momentum.  It may also be derived using the kinetic theory of
structureless molecules by starting with the Boltzmann transport
equation
\begin{equation}
\left(\frac{\partial}{\partial t} + \frac{p_i}{m} \frac{\partial}{\partial x_i}
+ F_i \frac{\partial}{\partial p_i}\right) f(r,p,t) = 
 \left[ \frac{\partial f(r,p,t)}{\partial t} \right]_{\mbox{\small \em
 coll}},
\end{equation}
where we follow the conventional definitions from Huang
\cite{Huang}. For the purpose of differentiating our approach to
arrive at post-NSE equations, we quickly comment on the assumptions of
the derivation.

First, collisional invariants $\chi$ are defined such that
\begin{equation}
\int \mathrm{d}^3 p\, \chi(r,p) \left[
\frac{\partial f(r,p,t)}{\partial t}
\right]_{\mbox{\small \em coll}} = 0.
\end{equation}
These collisional invariants are $\chi = m$ (mass),
$\chi = m v_i$ ($i = 1,2,3$ momentum), $\chi = \frac{1}{2}m|v -
u(r,t)|^2$ (thermal energy), where $u(r,t) = \langle v \rangle$.

To get the NSE, multiply the Boltzmann transport equation by $p$ and
integrate over all momentum, yielding
\begin{equation}
\rho\left(\frac{\partial}{\partial t} + u\cdot \nabla\right) u 
= \frac{\rho}{m} F - \nabla \cdot P
\end{equation}
where $\rho(r,t) = mn(r,t)$ and $P_{ij} = \rho\langle (v_i - u_i)(v_j
- u_j)\rangle$.  Our first comment is that the above equation results
from {\bf microscopic conservation laws assuming elastic collisions of
  point molecules}.

To arrive at the traditional NSE, we need an explicit form of the
pressure tensor, which is taken to be
\begin{equation}
P_{ij} = \delta_{ij} P - \mu \left[
\left( \frac{\partial u_i}{\partial x_j} + \frac{\partial u_j}{\partial x_i}\right)
- \frac{2}{3} \delta_{ij} \nabla \cdot u \right]
\end{equation}
where $\mu$ is the viscosity. The above choice of the pressure is
justified by the assumption that a fluid element, or microscopically,
as our second comment, {\bf the particle of the model has no intrinsic
angular momentum}.

In full component form, the Navier-Stokes equation is
\begin{eqnarray}
&&\rho\left(\frac{\partial}{\partial t} + u_j \frac{\partial}{partial x_j}\right) u_i
= \frac{\rho}{m} F_i \\*
&& - \frac{\partial}{\partial x_j} \left(
\delta_{ij} P - \mu \left[
\left( \frac{\partial u_i}{\partial x_j} + \frac{\partial u_j}{\partial x_i}\right)
- \frac{2}{3} \frac{\partial u_j}{\partial x_i} \nonumber
\right]\right). 
\end{eqnarray}

Notice that before the introduction of the definition of the pressure,
the conservation of momentum equation is exact in so far as the
Boltzmann transport equation is valid. This equation comes only from
the left-hand side of the Boltzmann equation. The contribution from
the collision term disappears by virtue of conservation of momentum
and the assumption of elastic collisions. Hence, only the left hand
side of the Boltzmann transport equation is important to yield the
conservation of momentum equation and NSE. We stress the two
assumptions needed to arrive at the Navier-Stokes equation: first,
elastic collisions, and second, the absence of angular momentum of the
structureless molecules.  What will happen if these two assumptions
are no longer valid?

\section{Correcting the Navier-Stokes Equation}
Suppose that each of the molecules could be found in any one of $N$
states, the ground state and $(N-1)$ excited states. Assume that
excitations and de-excitations are induced by molecular collisions,
which are now inelastic.  Then the semi-classical analogue of the
Boltzmann transport equation for each of $N$ distribution functions
will be
\begin{eqnarray}
\left(\frac{\partial}{\partial t} + \frac{p_i}{m} \frac{\partial}{\partial x_i}
+ F_i \frac{\partial}{\partial p_i}\right)f_n(r,p,t) =
&& \label{egtransport} \\*
\left[
\frac{\partial f_n(r,p,t)}{\partial t}\right]_{\mbox{\small \em inelastic}},
\nonumber
\end{eqnarray}
where the collision term might be replaced by
\begin{eqnarray}
\left[
\frac{\partial f_n(r,p,t)}{\partial t}\right]_{\mbox{\small \em inelastic}}
= \sum_{m\ne n}^N \gamma_{mn} J f_m(r,p) &&\\*
 - \sum_{m\ne n}^N
\gamma_{nm} J f_n(r,p) + \sigma K f_n(r,p) && \nonumber
\end{eqnarray}

$\gamma_{mn}$ is the transition probability of a particle in the $m$
state jumping to the $n$ state.  $J$ is a ``jump'' operator that will
carry the conservation law the we will invoke in the time evolution of
$N$ distribution probabilities.  $K$ is a ``kick'' operator that
allows the injection of energy from outside
\cite{Muriel1,Muriel2,Muriel3,Muriel4}.  $\sigma$ is the probability
that a particle is kicked to a different momentum by external
means. The kick operator makes the injection of energy into the system
possible. Eq. (\ref{egtransport}) is a generalization of our previous
models \cite{Muriel1,Muriel2,Muriel3,Muriel4}.

To calculate macroscopic averages, not only must we integrate over all
momentum, we should also sum over all $N$ states, to yield
\begin{equation}\label{epostnse}
\rho\left(\frac{\partial}{\partial t} + u\cdot\nabla\right)u = 
 \frac{\rho}{m} F - \nabla\cdot P + \mathit{driving}
+ \mathit{radiative},
\end{equation}
where
\begin{eqnarray}
&& \mathit{driving} = \sigma \sum_{n=1}^N \int
\mathrm{d}p^3 p K f_n(r,p), \label{edriving} \\
&& \mathit{radiative} = \label{eradiative} \\*
&&\sum_{n=1}^N \int \mathrm{d}p^3
p\left[ \sum_{m\ne n}^N \gamma_{mn} J f_m(r,p)
-\sum_{m\ne n}^N \gamma_{nm} J f_n(r,p) \right].\nonumber
\end{eqnarray}

We may think of (\ref{edriving}) and (\ref{eradiative}) as the quantum
corrections to the NSE.  The first sum (\ref{edriving}) is the driving
term.  We will call the second sum (\ref{eradiative}) the radiative
correction because every transition is accompanied by radiation. (NSE
ignores not only molecules, but also photons.)  The radiative term
represents the contribution of the internal degrees of freedom of
molecules to the macroscopic flow of a fluid. If all the transition
probabilities are zero, we simply reproduce the classical NSE. This
last equation, our post-NSE, no longer assumes elastic collisions and
the absence of angular momentum of the particles. The operators $J$
and $K$ have been defined in our earlier model calculations
\cite{Muriel1,Muriel2,Muriel3,Muriel4}, they may be redefined with new
models, but we will simplify them to show that even the simplest
application of Equation (\ref{epostnse}) leads to novel results.

We have had occasions to consider the radiative term
\cite{Muriel1,Muriel2,Muriel3,Muriel4}, and will consider them later
even more, but for now to arrive at immediate new results, we consider
only the influence of quantized kicks defined by the operator
$Kf_n(r,p) = f_n(r,p-\Pi)$ yielding
\begin{equation}
\int \mathrm{d}p^3 p K f_n(r,p) = 
mu + n(r,t) \Pi.
\end{equation}

\section{Examples}
Remove the force $F$ and drop the divergence of the pressure to obtain
a non-linear equation in one dimension
\begin{equation}
frac{\partial u}{\partial t} + \frac{1}{2} \frac{\partial u^2}{\partial x} -
\sigma u = \frac{\sigma \Pi}{m}.
\end{equation}
It is the simplest application of our post-NSE but which remains a
challenge still.

The stationary solution is to be obtained from 
\begin{equation}
\frac{1}{2} \frac{\partial u^2}{\partial x} - \sigma u = \frac{\sigma
  \Pi}{m},
\end{equation}
which is
\begin{equation}
u(x) = -\frac{\Pi}{m} \left[
W_k\left( -\frac{m}{\sigma \Pi} \mathrm{e}^{-\frac{m\sigma}{\Pi}
(x + C) - 1}\right) + 1 \right]
\end{equation}
where $C$ is a constant and $W_k$ denotes the
$k^\mathrm{th}$ branch of the Lambert W function.  The function
$W_k(z)$ is a solution of the equation $w \mathrm{e}^{w} = z$ in the
complex plane \cite{Corless}. {\bf The Lambert W function is
multi-valued, making the stationary average velocities multi-valued.}

\begin{figure}[h!]
\begin{center}
\begin{tabular}{cc}
Real part & Imaginary Part \\
\includegraphics[height=3cm]{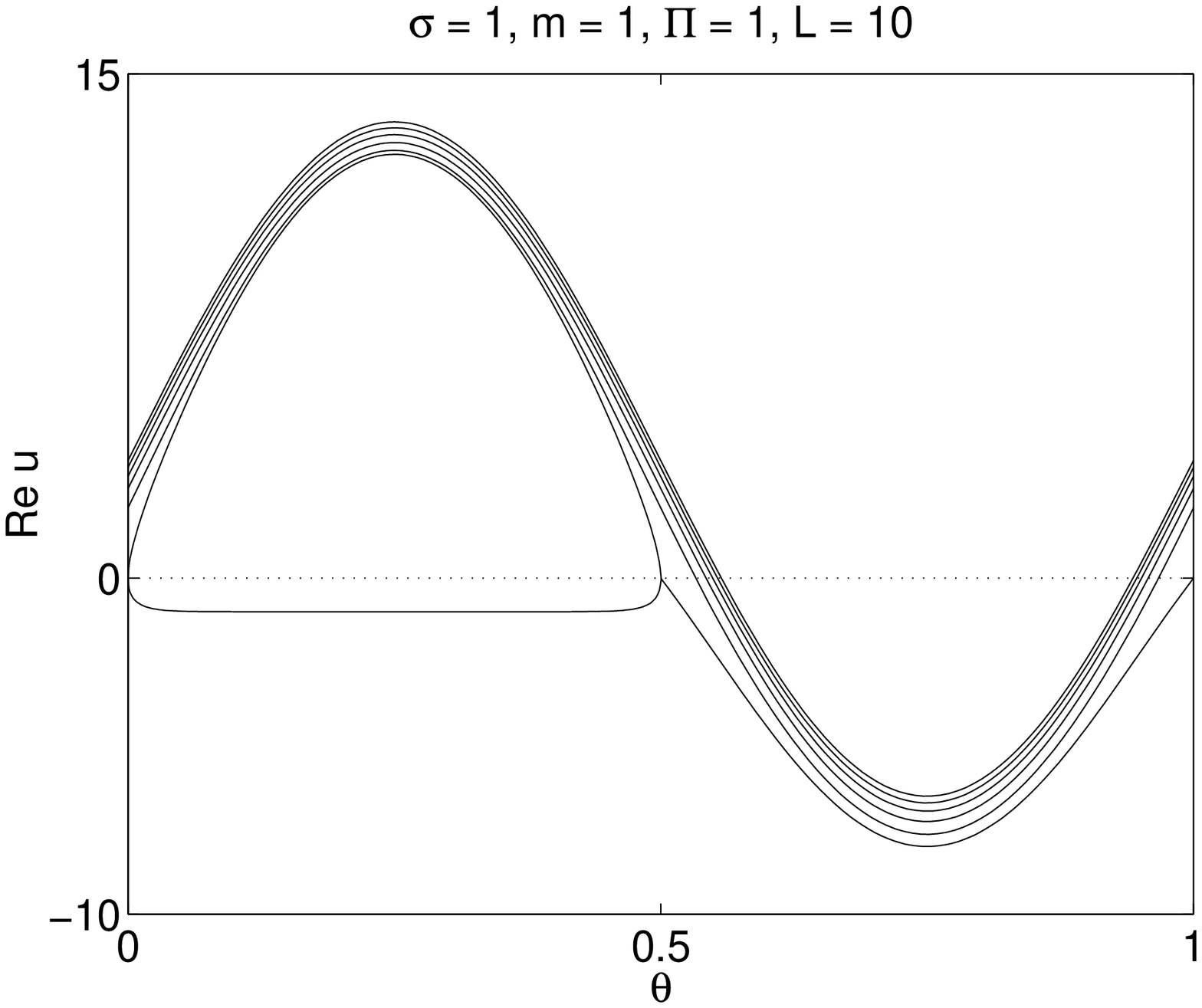} &
\includegraphics[height=3cm]{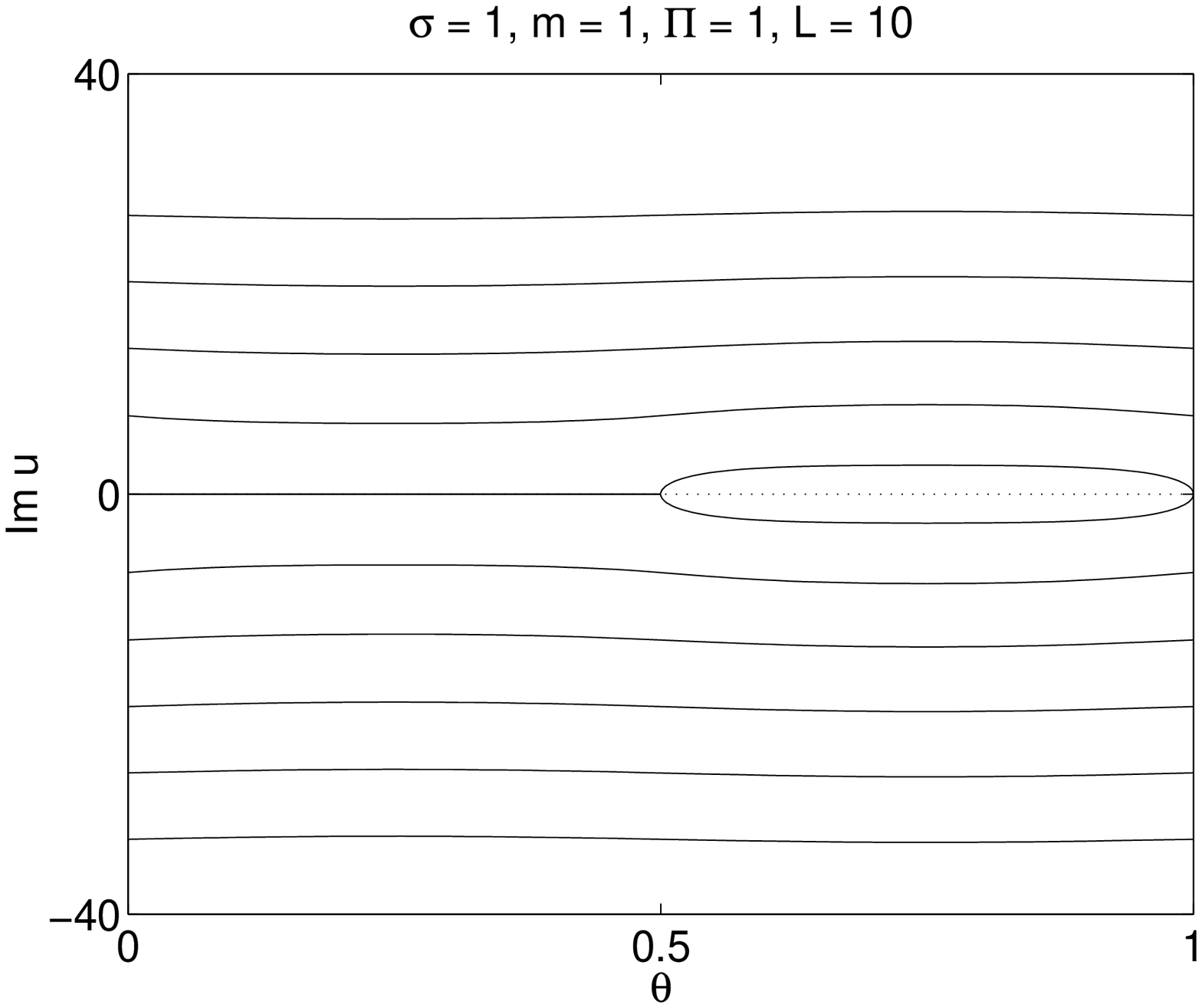} \\
\includegraphics[height=3cm]{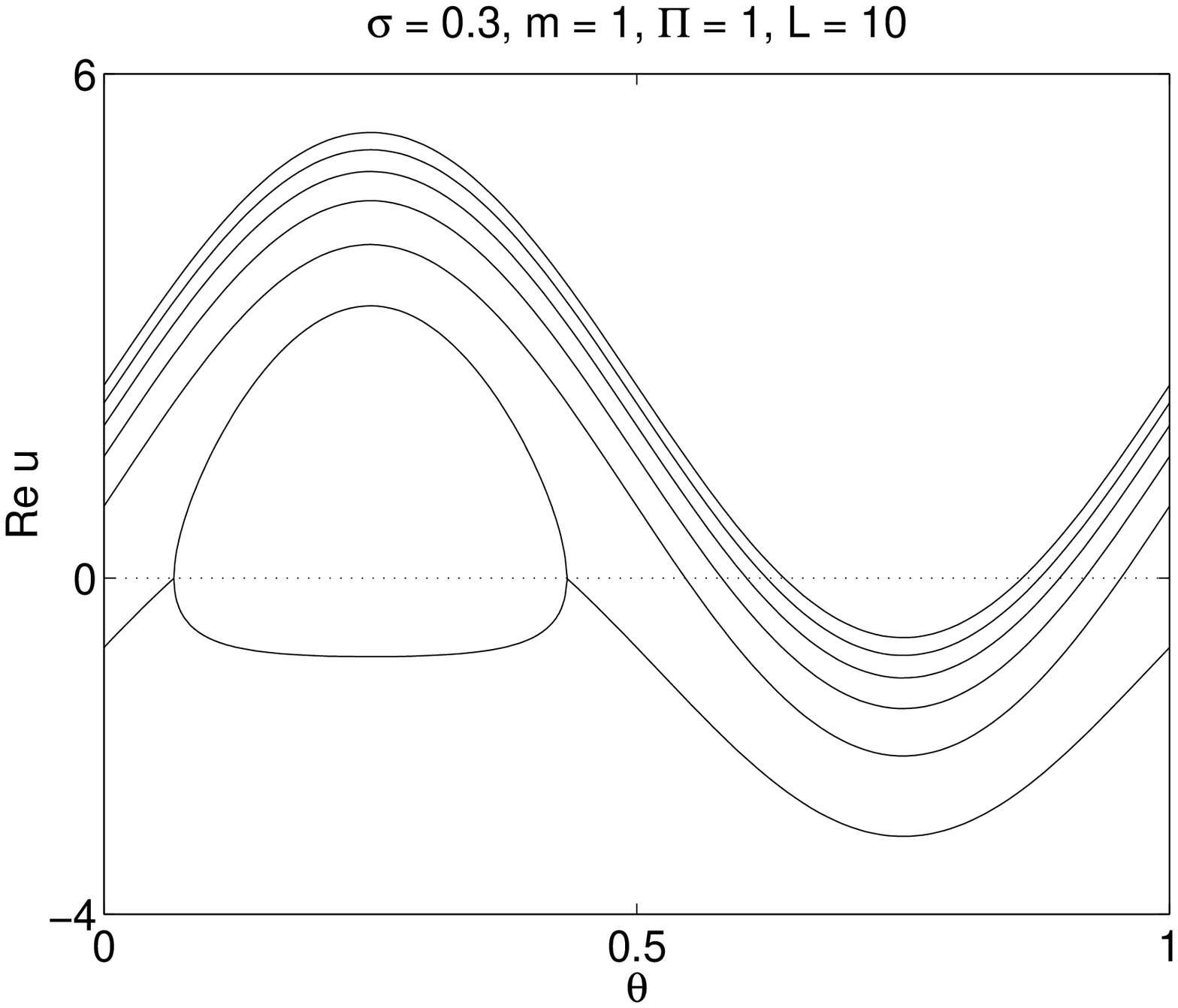} &
\includegraphics[height=3cm]{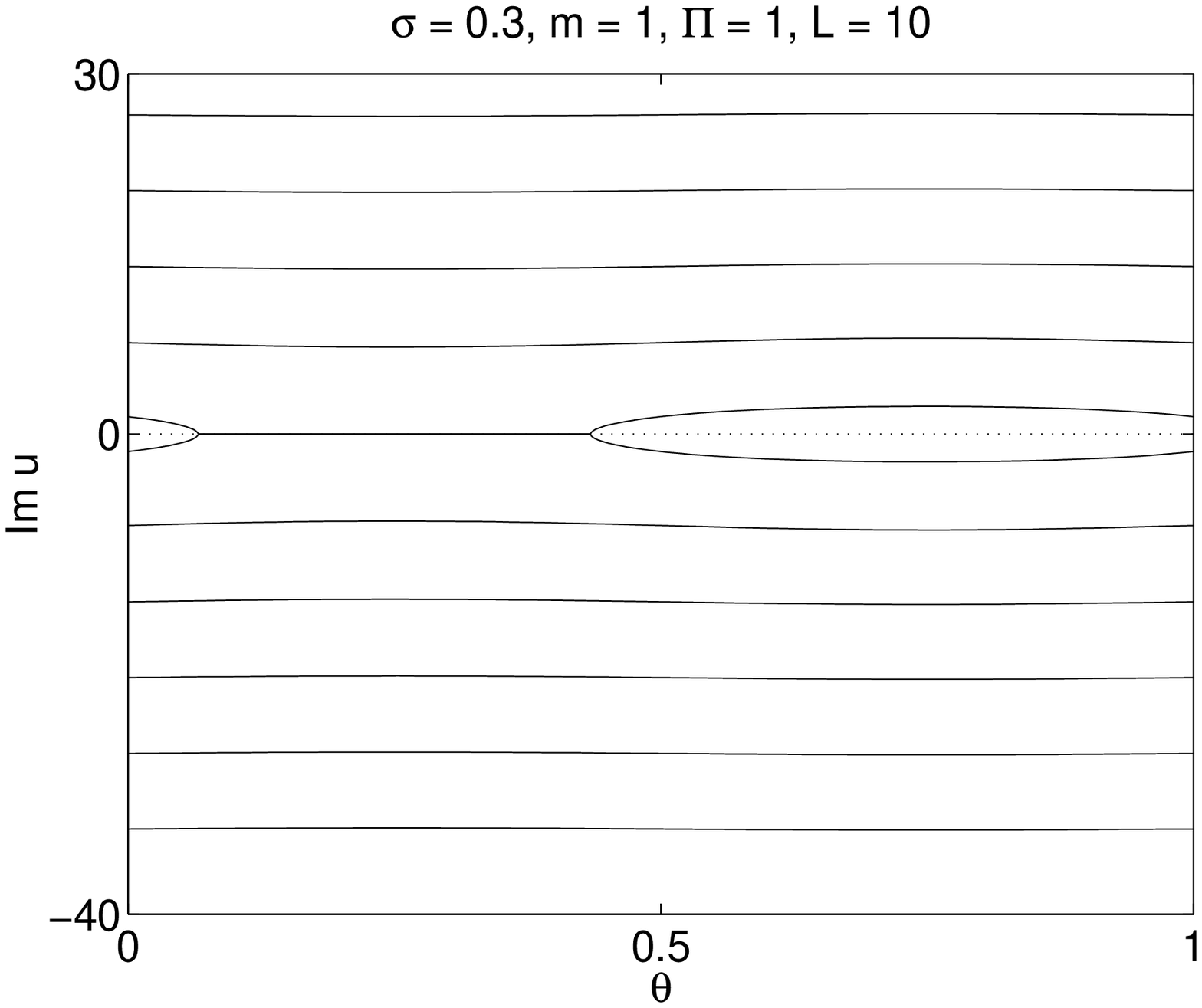} \\
\end{tabular}
\caption{\label{f1d}
Average velocity as a function of the angle for the torus
geometry. The principal branch is complemented by the other branches
which almost form a continuum.}
\end{center}
\end{figure}

We will choose a toroidal geometry, and put $x = L \sin(2\pi \theta)$,
$\theta = 0,\ldots 1$, to ensure periodic boundary conditions. The
physical model is one-dimensional, a donut of circumference $2\pi
L$. One could imagine a paddlewheel half-stuck into the donut to
provide quantum kicks to the fluid.  If $u(\theta=0) = 0$, then $C =
-\frac{\Pi}{\sigma m} \ln \Pi$ and we get
\begin{equation}
u_k(\theta) = -\frac{\Pi}{m} \left[
W_k\left(-\frac{m}{\sigma}
\ mathrm{e}^{-\frac{m\sigma}{\Pi}L \sin(2\pi\theta) - 1}
\right) + 1\right].
\end{equation}
We plot $u_k(\theta)$ in Figure \ref{f1d} for for $k = 0,\pm 1,\ldots \pm 5$.

Using the method of characteristics, we now find the
transient solutions of
\begin{equation}\label{etransient}
\frac{\partial u}{\partial t} + \frac{1}{2} \frac{\partial u^2}{\partial x} -
\sigma u = \frac{\sigma \Pi}{m}, \quad
u(x,0) = f(x),
\end{equation}
where $f(x)$ are the initial velocity averages.  Rewriting
(\ref{etransient}) as
$ u \frac{\partial u}{\partial x} + 
\frac{\partial u}{\partial t} = \sigma u + \frac{\sigma \Pi}{m}$, 
we identify its characteristic equations,
\begin{equation}\label{echar}
x'(\tau) = u(\tau), \quad t'(\tau) = 1, \quad
u'(\tau) = \sigma u(\tau) + \frac{\sigma \Pi}{m}. 
\end{equation}
The boundary conditions can be parameterized as
\begin{equation}
x_0(s) = s,\quad t_0(s) = 0, \quad
u_0(s) = f(s).
\end{equation}
For each fixed value of $s$, solving the characteristic equations (\ref{echar})
with initial values $x(0;s) = x_0(s)$, $t(0;s) = t_0(s)$, $u(0;s) =
u_0(s)$ yields a characteristic curve $u(x(\tau;s),t(\tau;s))$ in the
solution surface $u(x,t)$.  For more on the method of characteristics,
see for example \cite{Guenther}.

The solution $u(x,t)$ is
\begin{eqnarray}
u(x,t) &=& \mathrm{e}^{\sigma t}\left(
\frac{\Pi}{m} + f(s)\right) - \frac{\Pi}{m}, \\
x &=& s + f(s)\frac{\mathrm{e}^ {\sigma t} - 1}{\sigma}
+ \frac{\Pi}{m \sigma}(\mathrm{e}^{\sigma t} - \sigma t - 1),
\end{eqnarray}
where $s$ is defined implicitly by the second equation.  While it is
not in general possible to express $u(x,t)$ explicitly, we can still
interpret the solution in terms of the characteristic curves
$u(x(t;s),t)$.

\begin{figure}[h]
\begin{center}
\includegraphics[height=3cm]{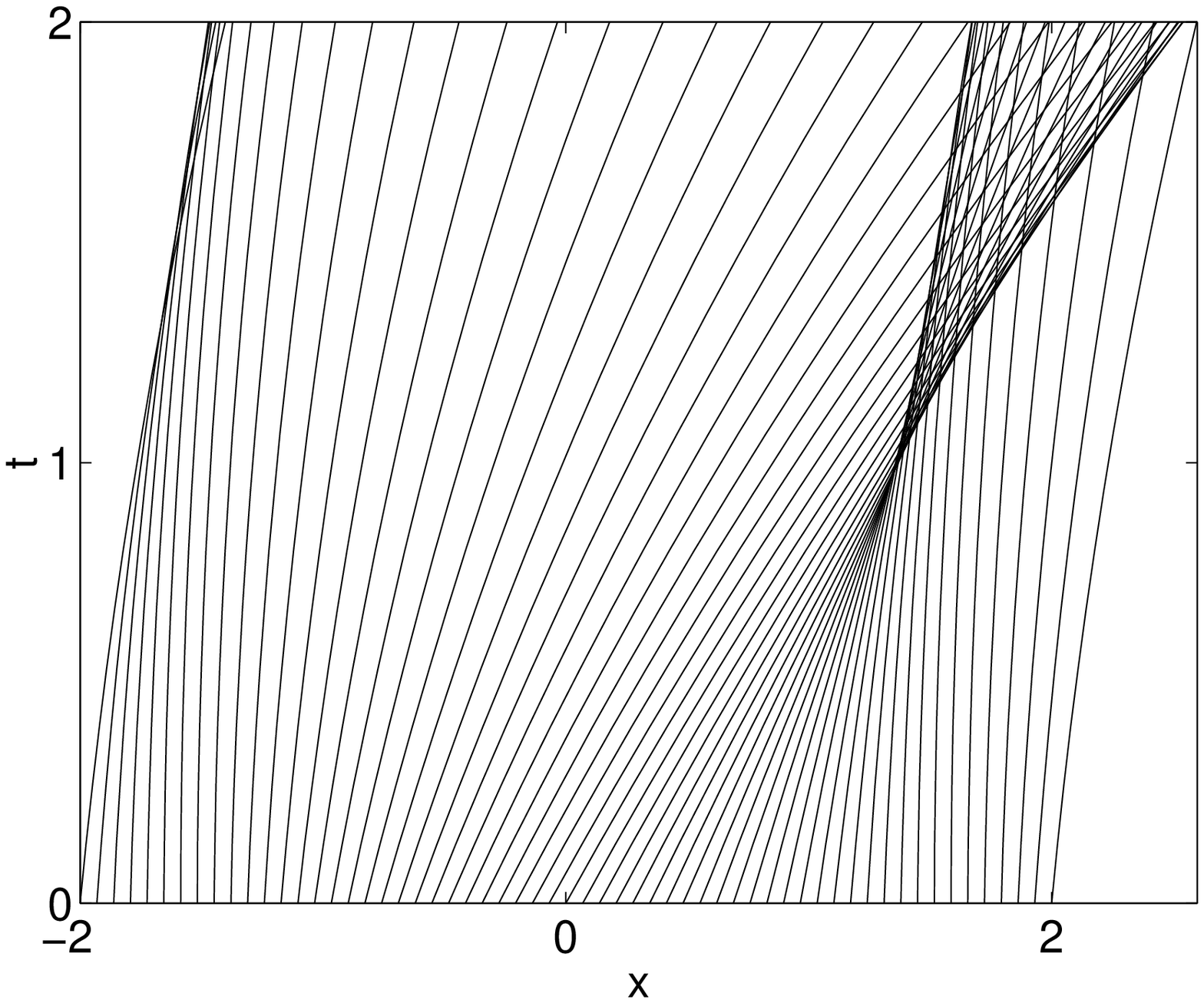} \qquad
\includegraphics[height=3cm]{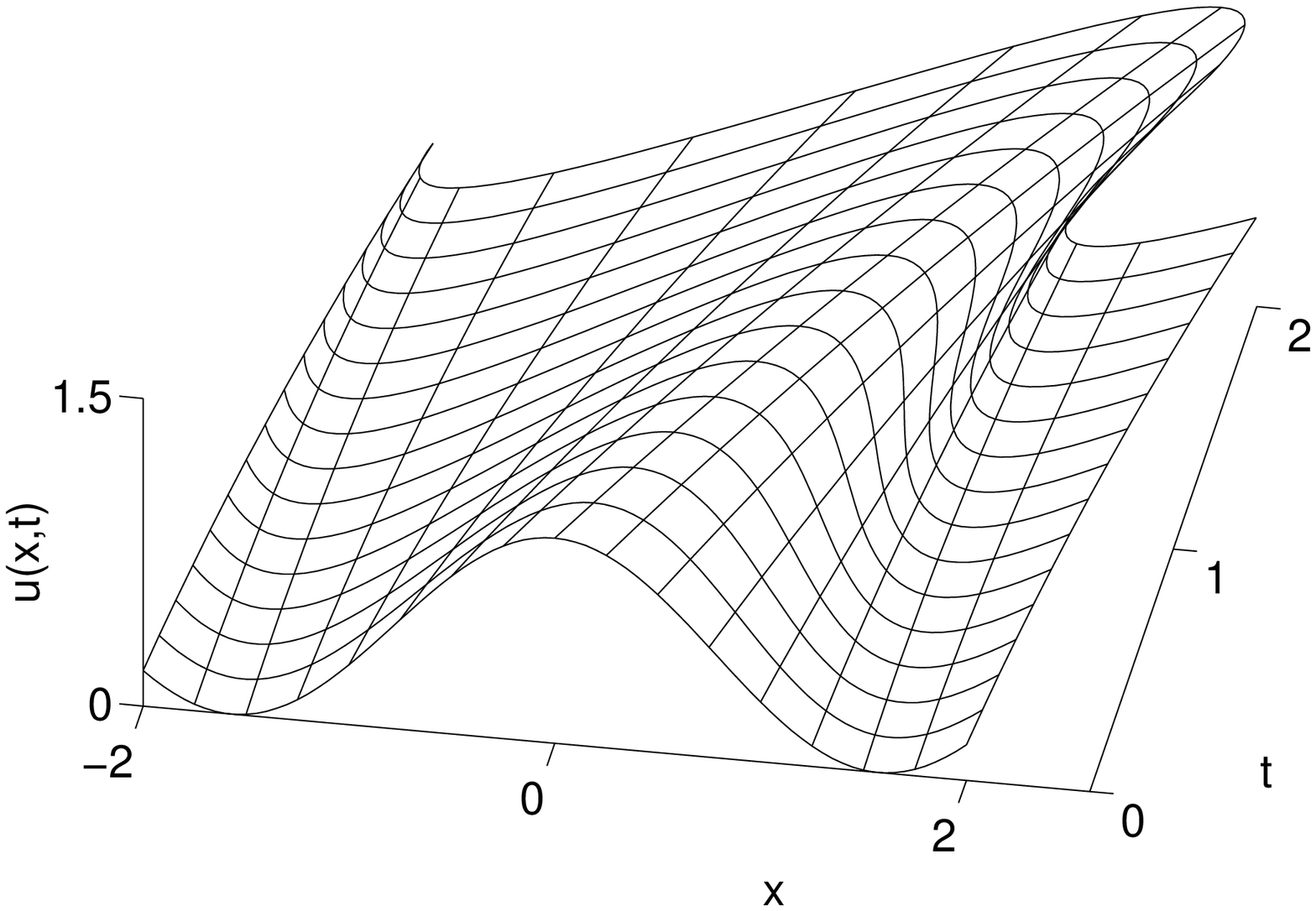} 
\caption{\label{ftrans} The transient problem with $\sigma = 0.1$, $m
= 1$, $\Pi = 1$, and $f(x) = \cos(x)^2$.  Left: The characteristics
cross at about $t = 1$.  Right: The solution surface $u(x,t)$ folds on
itself and becomes multi-valued where the characteristics cross.}
\end{center}
\end{figure}

Depending on the problem parameters $\sigma$, $m$, $\Pi$, and $f(x)$,
it is possible that the characteristic curves cross.  If the
characteristics cross at $(x_c,t)$, then there are multiple curves
$s_1, s_2, \ldots$ such that $x_c = x(t;s_1) = x(t;s_2) = \cdots$.
Moreover, the solution permits multiple values $u(x_c,t) =
u(x(t;s_1),t), u(x(t;s_2),t), \ldots$ at the crossing point.
Figure~\ref{ftrans} demonstrates this feature in a simple example.

\section{Conclusions}
It appears that in our first applications of a highly simplified
post-Navier-Stokes equation, we have arrived at multi-valued
velocities as a function of location. They may well be interpreted as
possible states of a turbulent system from which transitions to other
states may be possible. The possibility of velocity reversal, a
feature of turbulence, is immediately obvious. This result seems to be
the first instance of an analytic derivation of a multi-valued
velocity field and deserves further studies.

\bibliography{alpostnse}
\end{document}